\documentclass[aps, superscriptaddress, floatfix, reprint]{revtex4-1}
\usepackage{graphicx}
\usepackage{dcolumn}
\usepackage{bm}
\usepackage{amsfonts}
\usepackage{amsmath}
\usepackage{amssymb}
\usepackage{color}
\usepackage[normalem]{ulem}
\usepackage[percent]{overpic}

\usepackage[colorlinks=true,citecolor=blue]{hyperref}
\hypersetup{colorlinks=true,citecolor=blue,linkcolor=red,urlcolor=blue}

\def\be{\begin{equation}}
\def\ee{\end{equation}}
\def\rv{{\bf r}}
\def\kv{{\bf k}}

\begin{document}

\title{Thermoelectric response of nodal-line semimetals: probe of the Fermi surface topology}
\author{Shahin Barati}
\affiliation{Department of Physics, Institute for Advanced Studies in Basic Sciences (IASBS), Zanjan 45137-66731, Iran}
\affiliation{School of Nano Science, Institute for Research in Fundamental Sciences (IPM), Tehran 19395-5531, Iran}
\author{Saeed H. Abedinpour}
\email{abedinpour@iasbs.ac.ir}
\affiliation{Department of Physics, Institute for Advanced Studies in Basic Sciences (IASBS), Zanjan 45137-66731, Iran}
\affiliation{Research Center for Basic Sciences \& Modern Technologies (RBST), Institute for Advanced Studies in Basic Sciences (IASBS), Zanjan 45137-66731, Iran}
\affiliation{School of Nano Science, Institute for Research in Fundamental Sciences (IPM), Tehran 19395-5531, Iran}
\date{\today}

\begin{abstract}
We investigate the low-temperature thermoelectric properties of three-dimensional nodal-line semimetals within the semiclassical Boltzmann formalism. Considering short-range interaction between electrons and scattering agents, we calculate the anisotropic relaxation times and then obtain the charge conductivity and thermopower along the radial and the axial directions with respect to the nodal-line plane.  
Increasing the carrier concentration, a topological transition in the shape of the Fermi surface from a torus into an ellipsoid signal as a sharp change in the thermopower. An adequate treatment of the energy and direction dependence of the relaxation time is necessary for the observation of the topological transition of the Fermi surface in the thermoelectric properties. 
\end{abstract}
\maketitle

\section{Introduction}\label{introduction}

Topological semimetals have attracted great theoretical and experimental interest recently~\cite{hu2016evidence,borisenko2014experimental,lv2015experimental,liu2018experimental,burkov2011topological}. 
These materials generally include Weyl~\cite{balents2011weyl,burkov_prl2011, wan2011topological} or Dirac~\cite{wang2012dirac,Young, Zhijun} semimetals where the conduction and valence bands touch each other in a set of isolated points in the Brillouin zone, and the nodal-line semimetals~\cite{bzduvsek2016nodal,yu2017topological} where the touching between valence and conduction bands form open or closed lines in the Brillouin zone. The band touchings are protected by topological constraints.
Several interesting phenomena, such as large magnetoresistance, high bulk carrier mobility and quantum anomalous Hall effect~\cite{liang2015ultrahigh, zyuzin2012topological,lv2016extremely} are explored in topological semimetals. 

As the simplest form of the topological semimetals with closed nodal-lines, one can imagine a single circular band touching. Then the Fermi surface of an intrinsic system would be a circle, which transforms into a torus at low carrier doping and eventually forms an ellipsoid or more precisely a drum-head like surface with increasing the carrier concentration~\cite{li2018rules}.

Nodal-lines have been predicted in a large family of materials
~\cite{yu2015topological,kim2015dirac,xie2015new,weng2015topological,li2016dirac,lu2017two,xu_prb2017} and their existence has been experimentally verified in several components~\cite{bian2016topological,schoop2016dirac,neupane2016observation}. The quest for nodal-lines has even expanded to areas such as the ultra-cold atoms trapped in optical lattices~\cite{song2019observation} and electrical circuit lattices~\cite{luo2018topological}.

There has been a large body of studies on the charge transport properties of topological semimetals~\cite{ferrari2006raman, ashby2014chiral, tabert2016optical, barati2017optical}, and the thermoelectric properties of Weyl and Dirac semimetals have been also investigated~\cite{zhu2010universal, lundgren2014thermoelectric, mandal2018thermopower, salmankurt2017first}.

Thermoelectric materials are of great interest in the improvement of energy efficiency, as they can be used to harvest the waste heat~\cite{champier2017thermoelectric}. The main attention in the field of thermoelectric materials has been on heavily doped semiconductors, as their finite bandgap would result in the enhancement of their Seebeck coefficient leading to large figures or merit~\cite{Dehkordi2015mser}.
However, one of the main challenges of heavily doped semiconductors is their limited charge mobilities due to the impurity scatterings.

Very recently it has been proposed that, despite the common expectation, semimetals might be also very good candidates for thermoelectric applications~\cite{markov2018semi}. A large Seebeck coefficient in semimetals would result from the asymmetry between their valence and conduction bands.  The main advantages of semimetals over heavily doped semiconductors would be their large charge mobility in clean systems and also their low thermal conductivities if heavy mass elements are present among their constituents~\cite{markov2019thermoelectric}. 

In this work, we investigate the electronic contribution to the low-temperature thermoelectric properties of three-dimensional nodal-line semimetals. We have employed the semiclassical Boltzmann formalism and the generalized form of the relaxation time approximation which properly treats the asymmetry and energy dependence of the relaxation times. 
We are able to obtain fully analytic results for the charge and thermal conductivities as well as the thermopower of nodal-line semimetals in the presence of short-range scatterings. We also study the effect of approximating the relaxation time with an isotropic energy independent quantity, as such an approximation is very common in the ab-initio simulation of real materials. We observe that the low-temperature thermoelectric behavior of nodal-line semimetals is very sensitive to the details of the relaxation time.  We also suggest that the measurement of the thermopower is a simple yet very powerful tool to probe the topology of the Fermi surface in nodal-line semimetals. 

The rest of this paper is organized as follows. In Sec.~\ref{sec:model} we introduce our model low-energy Hamiltonian for the three-dimensional nodal-line semimetal and investigate its electronic dispersion and eigenstates. We have discussed the details of the semiclassical Boltzmann formalism for obtaining the thermoelectric responses and the method to calculate the anisotropic relaxation time in Sec.~\ref{sec:boltzmann}. Our analytic results for the low-temperature thermoelectric responses obtained from the constant and also from the anisotropic energy-dependent relaxation time approximations are presented in Sec.~\ref{sec:results}. Finally, we summarize our main findings in Sec.~\ref{sec:summary}.
We have also devoted Appendix~\ref{app:chemical} to the density of states and the chemical potential of a nodal-line semimetal and Appendix~\ref{app:relaxation} to the details of our analytical calculations for obtaining the anisotropic relaxation times.

\section{Model Hamiltonian for nodal-line semimetal}\label{sec:model}
\begin{figure}
\centering
\includegraphics[width=1\linewidth]{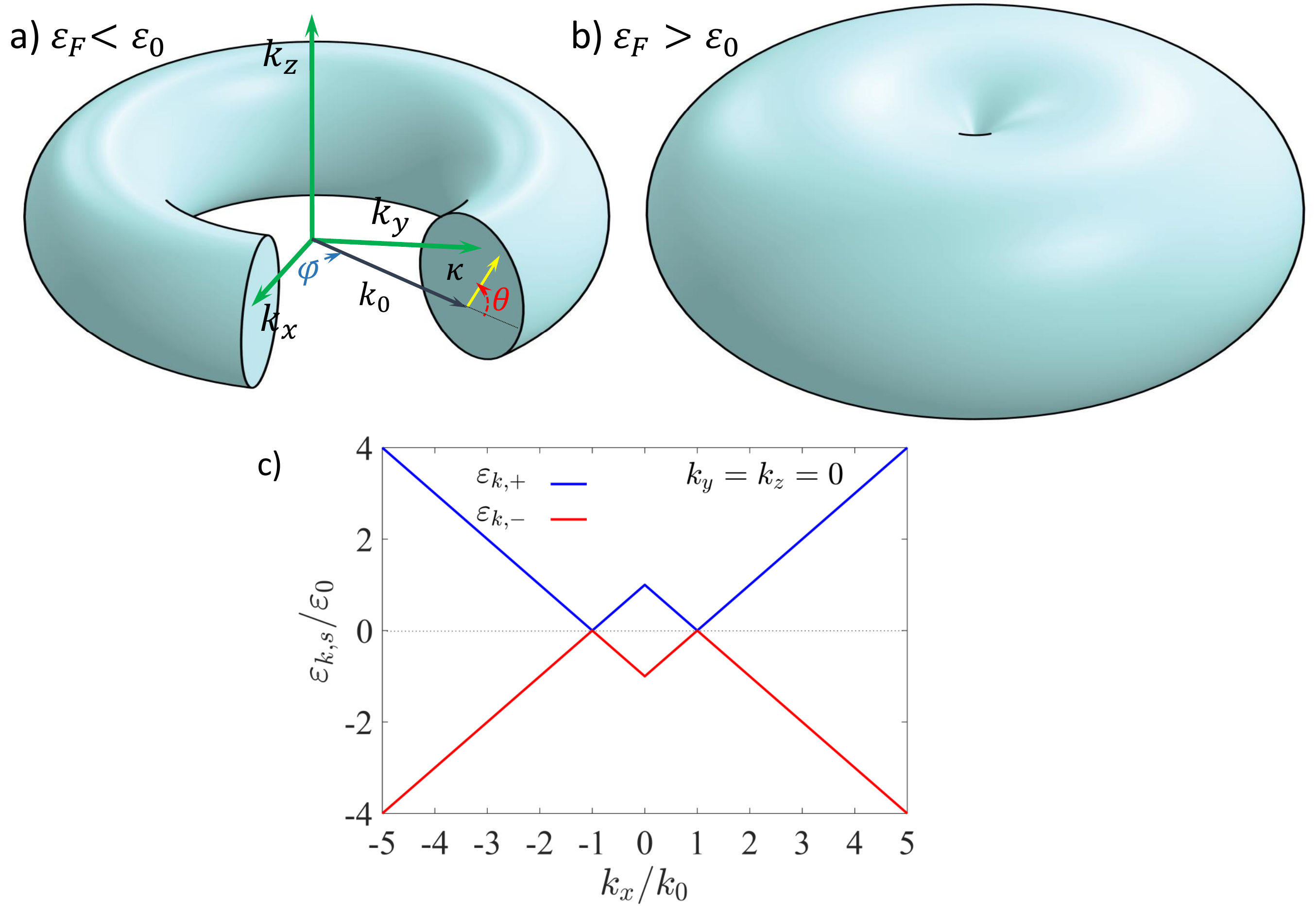} 
\caption{Sketches of the Fermi surface of a three dimensional nodal-line semimetal at two different Fermi energies $\varepsilon_{\rm F}<\varepsilon_0$ (a) and $\varepsilon_{\rm F}>\varepsilon_0$ (b). 
c) The low energy dispersions of electrons in a three dimensional nodal-line semimetal (in units of $\varepsilon_0$) versus $k_x/k_0$, for $k_y=k_z=0$.
\label{fig:Fermi}}
\end{figure}

In the continuum limit, the effective low-energy Hamiltonian~\cite{rhim2016anisotropic}
\be\label{eq:hamil1}
{\cal H}=\hbar \left( v k_\rho \tau_x + v_z k_z \tau_y\right),
\ee
gives a zero-energy nodal ring in the $x-y$ plane for $k_z=0$. 
Here, $\tau_{i}$ with $i=x,y$ are the Pauli matrices acting on the pseudo-spin (\emph{i.e.}, orbital) degree of freedom and $v$ and $v_z$ are the Fermi velocities in the radial and axial directions (with respect to the plane of the nodal line), respectively. 
In the cartesian coordinates we have $k_{\rho}=\sqrt{k^2_x+k^2_y}-k_0$, where $k_0$ is the radius of the nodal ring.
The eigen-energies of the Hamiltonian \eqref{eq:hamil1} are given by
\be \label{eq:energy}
\varepsilon_{\kv ,s} =s \hbar v \sqrt{k^2_{\rho}+\lambda^2 k^2_z},
\ee
where $s=+1 (-1)$ refers to the conduction (valance) band, and  $\lambda=v_z/v$. 
Note that even for $\lambda=1$ the system is anisotropic, due to the toroidal form of the constant-energy surface at low energies.
The Fermi surface evolves from a torus-like shape for $\varepsilon_{\rm F}< \varepsilon_0$ into a deformed ellipsoid for $\varepsilon_{\rm F} >\varepsilon_0$, where $\varepsilon_{\rm F}$ and $\varepsilon_0=\hbar v k_0$ are the Fermi energy and the characteristic energy of nodal ring, respectively. The energy dispersion and schematic sketches of the Fermi surface of Hamiltonian~\eqref{eq:hamil1} is illustrated in Fig.~\ref{fig:Fermi}.

With a transformation from the cartesian coordinates into the toroidal coordinates 
\be \label{eq:toroidal}
 \begin{aligned}
&k_x=\big(k_0+\kappa\cos\theta\big)\cos\phi,\\
&k_y=\big(k_0+\kappa\cos\theta\big)\sin\phi,\\
&k_z=\kappa\sin\theta /\lambda,
\end{aligned}
\ee
the Hamiltonian~\eqref{eq:hamil1} transforms to
\be\label{eq:hamil2}
{\cal H}=\hbar v\kappa\left(\cos\theta \tau_x +\sin\theta  \tau_y\right).
\ee
Note that, in the toroidal coordinates we have
$0 \leq \kappa \leq \infty$, $0\leq\phi \leq 2\pi$, and $-\pi+\theta_0 \leq \theta\leq \pi-\theta_0$, where $\theta_0=\arccos[{\rm max}(1,k_0/\kappa)]$ and the Jacobian determinant of this transformation is
$\mathcal{J}(\kappa,\theta,\phi)=\kappa\left(k_0+\kappa\cos\theta\right)/\lambda$.
The eigenvalues of Hamiltonian~\eqref{eq:hamil2} take the simple form $\varepsilon_{\kv ,s}=s\hbar v \kappa$ and their corresponding eigenstates read
\be\label{eq:eigenstate}
\psi_{\kv,s}(\rv)
=\dfrac{1}{\sqrt{2V}}
\left(
\begin{array}{c}
1\\
s e^{i\theta} \\
\end{array} \right) e^{i\kv\begin{tiny}
\cdot
\end{tiny}\rv},
\ee
with $V$ the sample volume. 

Finally, the cartesian components of the group velocity of electrons ${\bf v}_{\kv,s}=\nabla_{\kv} \varepsilon_{\kv,s}/\hbar$, read
\be \label{eq:velocity}
\begin{aligned}
v^x_{\kv,s}&=s v\cos\theta\cos\phi,\\
v^y_{\kv,s}&=s v\cos\theta\sin\phi,\\
v^z_{\kv,s}&=s \lambda v\sin\theta.
\end{aligned}
\ee

\section{Semiclassical theory of thermoelectric response}\label{sec:boltzmann}
To investigate the thermoelectric properties of a nodal-line semimetal, we assume that the temperature is low enough, such that only the electronic degrees of freedom contribute to the thermoelectric coefficients. 
In the linear response regime, the charge ${\bf j}$ and thermal ${\bf j}^Q$ current densities in response to an external electric field ${\bf E}$ and temperature gradient ${\bm \nabla} T$, can be written as~\cite{ashcroftsolid}
\be
\begin{aligned}
&j_{\alpha}=L^{11}_{\alpha\beta}E_{\beta}+L^{12}_{\alpha\beta}(-{\bm \nabla} T)_{\beta},\\
&j^{Q}_{\alpha}=L^{21}_{\alpha\beta}E_{\beta}+L^{22}_{\alpha\beta}(-{\bm \nabla} T)_{\beta},
\end{aligned}
\ee
where $\alpha$ and $\beta$ refer to three spatial directions $x, y,$ and $z$, the set of $L^{ij}$ tensors are the thermoelectric coefficients, and the summation over repeated indices is implied.
Theses thermoelectric coefficients at temperature $T$ are given by
$L^{11}_{\alpha\beta}=\mathcal{L}^0_{\alpha\beta}$, $L^{21}_{\alpha\beta}=TL^{12}_{\alpha\beta}=-\mathcal{L}^1_{\alpha\beta}/e$, and $L^{22}_{\alpha\beta}=\mathcal{L}^2_{\alpha\beta}/(e^2T)$,
where $-e$ is the charge of electron and 
\be \label{eq:transport1}
\mathcal{L}^n_{\alpha\beta}=\int \mathrm{d}\varepsilon \left(\varepsilon-\mu\right)^n\sigma_{\alpha\beta}(\varepsilon)\left[-\dfrac{\partial f(\varepsilon)}{\partial \varepsilon}\right].
\ee
Here, $\mu$ is the chemical potential, $f(\varepsilon)=1/[e^{(\varepsilon-\mu)/(k_{\rm B}T)}+1]$ is the equilibrium Fermi distribution function, and the generalized transport distribution function is defined as~\cite{gorsso_book}
\be \label{eq:sigma}
\sigma_{\alpha\beta}(\varepsilon)=e^2 g \sum_s\int \dfrac{d^3k}{(2\pi)^3}\delta(\varepsilon-\varepsilon_{\kv,s})v^{\alpha}_{\kv,s}v^{\beta}_{\kv,s}\tau^{\beta}_{\kv,s},
\ee
where $g$ is the degeneracy factor and $\tau^{\beta}_{\kv,s}$ is the momentum relaxation-time of electrons along the  $\beta$ direction, which will be discussed in details later in this section.

In the absence of an external magnetic field, the off-diagonal components of the thermoelectric responses vanish. Furthermore, the toroidal symmetry of the Hamiltonian implies $L^{ij}_{xx}=L^{ij}_{yy}$. Therefore we only investigate the diagonal thermoelectric responses along the axial (\emph{i.e.}, $z$) and radial (\emph{i.e.}, $x$) directions.

To obtain the thermoelectric coefficients, one needs to obtain the generalized transport distribution function $\sigma_{\alpha\alpha}(\varepsilon)$, and then use Eq.~\eqref{eq:transport1} to find all other thermoelectric coefficients. 
In the following, without loss of the generality, we will consider only electron-doped systems \emph{i.e.}, $\varepsilon_{\rm F}>0$. In this case, only the conduction band will contribute to the thermoelectric response at low temperatures $k_{\rm B}T << \varepsilon_{\rm F}$. Therefore, we will drop the band index in the following for the notational brevity. 

If the transport distribution function $\sigma_{\alpha\alpha}(\varepsilon)$ is a smooth and differentiable function of energy, which is valid away from the nodal ring energy $\varepsilon_0$, with the help of the Sommerfeld expansion to the leading-orders in temperature, we obtain~\cite{gorsso_book}
\be \label{eq:lowT}
\begin{aligned}
&\mathcal{L}^0_{\alpha\alpha}\approx \sigma_{\alpha\alpha}(\varepsilon_{\rm F}),\\
&\mathcal{L}^1_{\alpha\alpha}\approx\frac{\pi^2}{3}(k_{\rm B}T)^2 \sigma'_{\alpha\alpha}(\varepsilon_{\rm F}),\\
&\mathcal{L}^2_{\alpha\alpha}\approx\frac{\pi^2}{3}(k_{\rm B}T)^2 \sigma_{\alpha\alpha}(\varepsilon_{\rm F}),\\
\end{aligned}
\ee
where $\sigma'$ is the derivative of the generalized transport distribution function with respect to energy.
Note that $\sigma_{\alpha\alpha}(\varepsilon_{\rm F})$ is indeed the zero temperature charge conductivity.
From Eqs.~\eqref{eq:lowT} we immediately recover the Mott formula for the Seebeck coefficient (thermopower)
\be\label{eq:Mott}
S_{\alpha\alpha}=\frac{L^{21}_{\alpha\alpha}}{T L^{11}_{\alpha\alpha}}
=-\frac{\pi^2 k_{\rm B}^2 T}{3e} \frac{\sigma'_{\alpha\alpha}}{\sigma_{\alpha\alpha}},
\ee
and the Wiedemann-Franz law for the electronic thermal conductivity
\be\label{eq:WF}
\kappa^e_{\alpha\alpha}=L^{22}_{\alpha\alpha}-\frac{L^{21}_{\alpha\alpha}L^{12}_{\alpha\alpha}}{L^{11}_{\alpha\alpha}}
=L T \sigma_{\alpha\alpha},
\ee
with $L= \pi^2 k_{\rm B}^2 /(3 e^2)$ the Lorentz number. Here, for brevity we have omitted all the arguments which are the Fermi energy. 
One of the characteristics of thermoelectric materials is their potential for energy conversion, and thus these materials could be used as thermoelectric generators. The key parameter that defines the efficiency of a thermoelectric generator is characterized by the dimensionless thermoelectric figure of merit ${\rm zT}\equiv S^2\sigma T/\kappa$, 
where $\kappa$ is the total, the sum of electronic and lattice contributions to the thermal conductivity. As we are investigating only the electronic degrees of freedom here, we can only estimate the upper bound for the figure of merit. 
Note that these relations are valid at low temperatures $k_{\rm B} T \ll \varepsilon_{\rm F}$, and for Fermi energies away form the nodal ring energy, i.e. $k_{\rm B} T \ll |\varepsilon_{\rm F}-\varepsilon_0|$. 

\subsection{Relaxation times}\label{sec:relaxation}
The momentum relaxation time of electrons which appears in the definition of the thermoelectric coefficients, in general, depends on the scattering mechanism, details of the electronic band structure, and also the wave function of electrons. In principle, each of these three factors can make transport in a medium anisotropic~\cite{titf_jpcm2015, zarezad_prb2018,trushin_prb2019}.
The simplest approximation, however, is to take the relaxation time $\tau$ a constant parameter. 
This approximation is usually adopted in many numerical packages, such as the BoltzTraP~\cite{BoltzTraP}, therefore only the contributions from the band structure, as obtained for example from the ab-initio calculations, are accounted for in the thermoelectric properties.

For an anisotropic system, the relaxation time depends on both the magnitude and the direction of $\kv$, and could be obtained from the integral equation~\cite{kim2019vertex}
\be \label{eq:relaxtimeaniso}
1=\int \dfrac{\mathrm{d}^3 k'}{(2\pi)^3}W_{\kv\kv'}\left(\tau^{\alpha}_{\kv}-\dfrac{v^{\alpha}_{\kv'}}{v^{\alpha}_{\kv}}\tau^{\alpha}_{\kv'}\right),
\ee
where $W_{\kv\kv'}$ is the transition rate between two eigenstates $\kv$ and $\kv'$ of the system. 
Within the first Born approximation, for elastic scatterings and uncorrelated disorders, we find the Fermi's golden rule 
 \be
W_{\kv\kv'}=\dfrac{2\pi}{\hbar}n_{\rm imp}|V_{\kv\kv'}|^2\delta(\varepsilon_{\kv}-\varepsilon_{\kv'}).
\ee
Here, $n_{\rm imp}$ is the density of impurities in the sample, and $V_{\kv\kv'}$ is the impurity potential describing a scattering of electron from $\kv$ to $\kv'$.
For an isotropic system, Eq.~\eqref{eq:relaxtimeaniso} simplifies to the familiar textbook expression
$1/\tau_{k}=\int \mathrm{d}^3k' W_{\kv\kv'}(1-\cos\theta_{\kv\kv'})/(2\pi)^3$,
where the relaxation time depends only on the magnitude of the wave vector~\cite{ashcroftsolid}.

In this work, we obtain analytic results for the anisotropic relaxation time and then the thermoelectric coefficients of a three-dimensional nodal-line semimetal in the presence of isotopic short-range impurity scatterings.
For the sake of completeness, we compare our results with the ones obtained within the constant relaxation time approximation. 

\section{Analytic Results for the Thermoelectric Responses}\label{sec:results}
In this section, we present our results for the low-temperature thermoelectric responses of a three-dimensional nodal line semimetal. Before discussing our main results obtained from the anisotropic relaxation time given by Eq.~\eqref{eq:relaxtimeaniso}, we show what one would get from the constant relaxation time approximation.

\subsection{Thermoelectric responses with constant relaxation time}
\begin{figure}
\centering
\includegraphics[width=\linewidth]{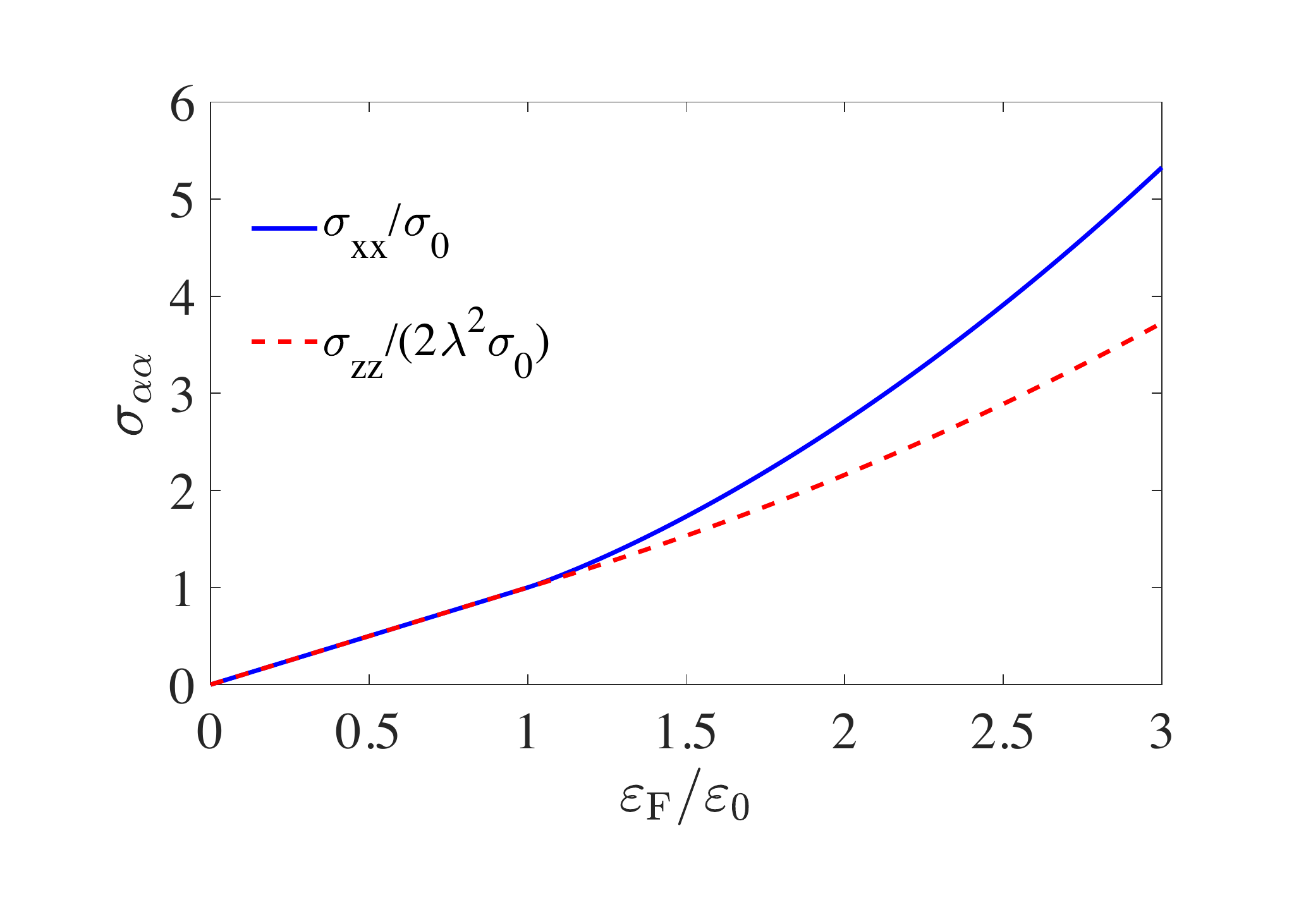}\\
\caption{The zero temperature static charge conductivities versus the Fermi energy $\varepsilon_{\rm F}$ along the axial (solid blue) and axial (dashed red) directions (in units of $\sigma_0$ and $2\lambda^2\sigma_0$, respectively), obtained within the constant relaxation time approximation.
\label{fig:sigmaz-xtc}}
\end{figure}

\begin{figure}
\centering
\includegraphics[width=1\linewidth]{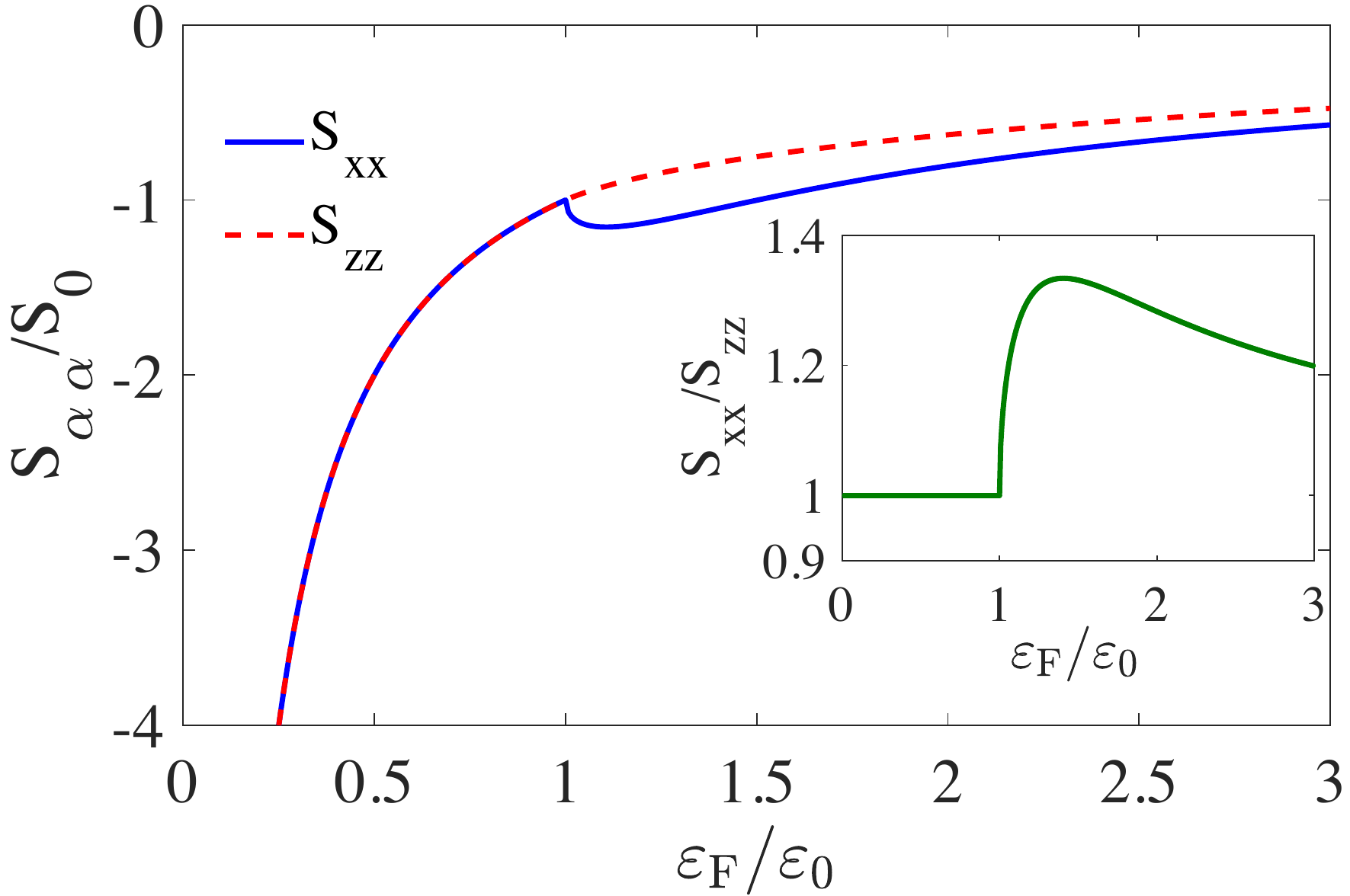}
\caption{The Fermi energy dependance of the radial (solid blue) and axial (dashed red) thermopowers (in units of $S_0$) obtained using the constant relaxation time. 
Inset: The ratio between radial and axial thermopowers $S_{xx}/S_{zz}$ versus the Fermi energy. 
\label{fig:seebeckh}}
\end{figure}

We start with replacing the relaxation time $\tau^\beta_\kv$ in Eq.~\eqref{eq:sigma} with a momentum and direction independent parameter $\tau_0$. 
the integral over $\kv$ is analytically solvable for both radial and axial directions, resulting in
\be\label{eq:sigma_iso}
\begin{aligned}
\sigma_{xx}(\varepsilon)&=\sigma_0{\tilde \varepsilon}\left[1+\Theta(\tilde \varepsilon-1) f_{x}(\tilde \varepsilon)\right]\\ 
\sigma_{zz}(\varepsilon)&=\sigma_0 2 \lambda^2{\tilde \varepsilon}\left[1+\Theta(\tilde \varepsilon-1)f_z(\tilde \varepsilon)\right], 
\end{aligned}
\ee
for the generalized transport distribution function, with $\sigma_0=(g e^2 \tau_0 k^2_0 v)/(8\pi \hbar \lambda)$, $\tilde \varepsilon=\varepsilon/\varepsilon_0$, and 
\be
\begin{aligned}
f_x(x)&=\frac{1}{\pi}\left[\left(\frac{4x^2-1}{3x^2}\right)\sqrt{x^2-1}-{\rm arcsec}(x)\right],\\
f_z(x)&=\frac{1}{\pi}\left[\left(\frac{2x^2+1}{3x^2}\right)\sqrt{x^2-1}-{\rm arcsec}(x)\right].
\end{aligned}
\ee
For the charge conductivity at zero temperature one only needs to replace the energy $\varepsilon$ on the right-hand-side of Eq.~\eqref{eq:sigma_iso} with the Fermi energy $\varepsilon_{\rm F}$.
In Fig.~\ref{fig:sigmaz-xtc} we have illustrated the Fermi energy dependance of conductivities within the constant relaxation time approximation.
Constant relaxation time approximation predicts linear dependance of both components of the charge conductivity on the Fermi energy at low carrier concentrations. The conductivity vanishes at the charge neutrality point i.e., $\varepsilon_{\rm F}=0$. Finally, the axial and radial components of the charge conductivity are different, even for the most symmetric case, i.e., $\lambda=1$ where the axial conductivity is twice the radial one at low Fermi energies. 
The finite temperature thermoelectric coefficients could be readily obtained with the help of Eq.~\eqref{eq:lowT}. In particular, for the thermopower we find 
\be\label{eq:s_iso}
S_{\alpha\alpha}\left(\varepsilon_{\rm F}\right)=-\frac{S_0}{\tilde \varepsilon_{\rm F}}
\left[1+\Theta(\tilde \varepsilon_{\rm F}-1)\frac{g_{\alpha}(\tilde \varepsilon_{\rm F})}{1+f_{\alpha}(\tilde \varepsilon_{\rm F})}\right],
\ee
where $S_0=\pi^2 k_{\rm B}^2 T/(3 e \varepsilon_0)$, $\tilde \varepsilon_{\rm F}=\varepsilon_{\rm F}/\varepsilon_0$, and
\be
\begin{aligned}
g_x(x)&=\frac{2(2x^2+1)\sqrt{x^2-1}}{3\pi x^2},\\
g_z(x)&=\frac{2\sqrt{(x^2-1)^3}}{3\pi x^2}.
\end{aligned}
\ee
Note that, for $\varepsilon_{\rm F}<\varepsilon_0$, we find $S_{\alpha\alpha}=-\pi^2k_{\rm B}^2 T/(3 e\varepsilon_{\rm F})$
which is direction independent.
The behavior of thermopower in different directions versus the Fermi energy, as well as the ratio between the axial and radial thermopowers within the constant relaxation time approximation, is illustrated in Fig.~\ref{fig:seebeckh}. The thermopower
is isotropic for $\varepsilon_{\rm F}<\varepsilon_0$ and becomes anisotropic at higher energies. It is interesting to notice that the results do not depend directly on the ratio between two Fermi velocities in different directions $\lambda$, and the maximum anisotropy is observed for $\varepsilon_{\rm F} \approx 1.5 \,\varepsilon_0$. 

\subsection{Thermoelectric responses with anisotropic relaxation times}
As the electronic structure of a three-dimensional nodal-line semimetal is anisotropic, one should employ the anisotropic relaxation time obtained from Eq.~\eqref{eq:relaxtimeaniso} to investigate the thermoelectric properties. 
Here we consider scatterings from short-range impurities i.e., $V_{\kv,\kv'}=V_0$, therefore the transition rate between two states reads
\be
W_{\kv\kv'}=u_0 \left[1+\cos\left(\theta -\theta'\right)\right]\delta(\kappa - \kappa'),
\ee
where $u_0=\pi n_{\rm imp}V^2_0/(\hbar^2 v)$. 
Inserting $W_{\kv\kv'}$ into Eq. \eqref{eq:relaxtimeaniso}, after some lengthly algebra (see, Appendix~\ref{app:relaxation} for details), we obtain
\be\label{eq:tau_aniso}
\begin{aligned}
\tau^{x}_{\kv}&=\frac{\tau_0}{\left(b_0+b_c\cos\theta\right)\tilde \kappa}+...,\\
\tau^{z}_{\kv}&=\frac{\tau_0}{\left(b_0+b_c\cos\theta\right)\tilde \kappa}\frac{1}{1-\gamma_s}+...,
\end{aligned}
\ee
where $\tilde \kappa=\kappa/k_0$, $\tau_0=4\pi \lambda/(u_0 k_0^2)$, and the parameters $b_i$ and $\gamma_i$ are defined in Appendix~\ref{app:relaxation}. 
The omitted terms on the right-hand side of Eq.~\eqref{eq:tau_aniso} do not survive the angular integrations in Eq.~\eqref{eq:sigma} (see, the Appendix~\ref{app:relaxation} for details) and therefore do not contribute to the thermoelectric coefficients. 
The generalized transport distribution functions read
\be\label{eq:sigma_aniso}
\begin{aligned}
&\sigma_{xx}(\varepsilon)=\sigma_0\gamma_c(\tilde \varepsilon),\\
&\sigma_{zz}(\varepsilon)=\sigma_02\lambda^2\frac{\gamma_s(\tilde \varepsilon)}{1-\gamma_s(\tilde \varepsilon)},
\end{aligned}
\ee
with $\sigma_0=(g e^2\tau_0 k^2_0 v)/(8 \pi \hbar \lambda)=(g e^2 v)/(2\hbar u_0)$, which is defined in a similar fashion to the constant relaxation time approximation case but the explicit expression for $\tau_0$, given right after Eq.~\eqref{eq:tau_aniso}, is replaced therein. 
For $\varepsilon<\varepsilon_0$, these expressions simplify to
\be\label{eq:sigma_aniso_low}
\begin{aligned}
&\sigma_{xx}(\varepsilon<\varepsilon_0)= \sigma_0 \left[1+\frac{4}{\tilde{\varepsilon}^2}\left(1-\dfrac{2}{\sqrt{4-\tilde{\varepsilon}^2}}\right)\right],\\
&\sigma_{zz}(\varepsilon<\varepsilon_0)=\sigma_0 \lambda^2\sqrt{4-\tilde{\varepsilon}^2}.
\end{aligned}
\ee
We recall that the zero-temperature conductivities are obtained after replacing $\varepsilon$ with Fermi energy on the right hand side of Eq.~\eqref{eq:sigma_aniso}. 

\begin{figure}
\centering
\includegraphics[width=0.99\linewidth]{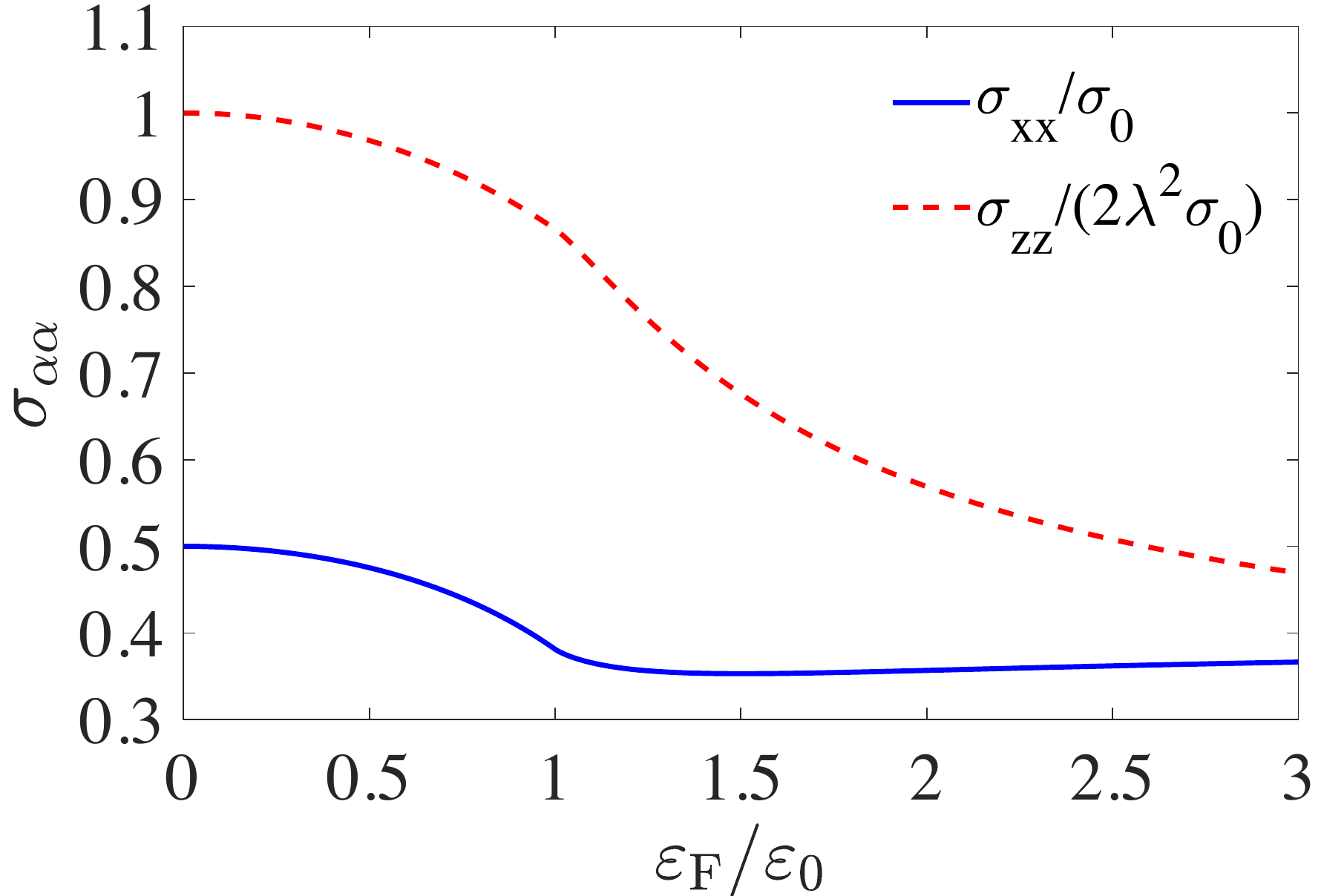}
\caption{The dependence of radial (solid blue) and axial (dashed red) static charge conductivities  (in units of $\sigma_0$ and $2\lambda^2\sigma_0$, respectively) on the Fermi energy $\varepsilon_{\rm F}$ at zero temperature obtained from the anisotropic relaxation time.
\label{fig:sigma}}
\end{figure}

In Fig.~\ref{fig:sigma} we have illustrated the Fermi energy dependence of the static charge conductivities.
They are several fundamental differences in comparison with the results obtained within the constant relaxation time approximation. First of all, the radial conductivity is independent of the anisotropy factor $\lambda$, as $\sigma_0$ does not depend on it in the anisotropic case. Second, in contrast to the constant relaxation time case, the intrinsic (i.e., $\varepsilon_{\rm F}=0$) conductivities are non-zero. More explicitly, we have 
$\sigma_{xx}(0)= \sigma_0/2$ and $\sigma_{zz}(0)= 2\lambda^2\sigma_0$, and finally the conductivities at high carrier concentration weekly depend on the Fermi energies.

The electronic thermal conductivity is related to the charge conductivity through the Wiedemann–Franz law at low temperatures, so its anisotropy and Fermi energy dependance are identical to what we already discussed about $\sigma_{\alpha\alpha}$.

\begin{figure}
\centering
\includegraphics[width=1\linewidth]{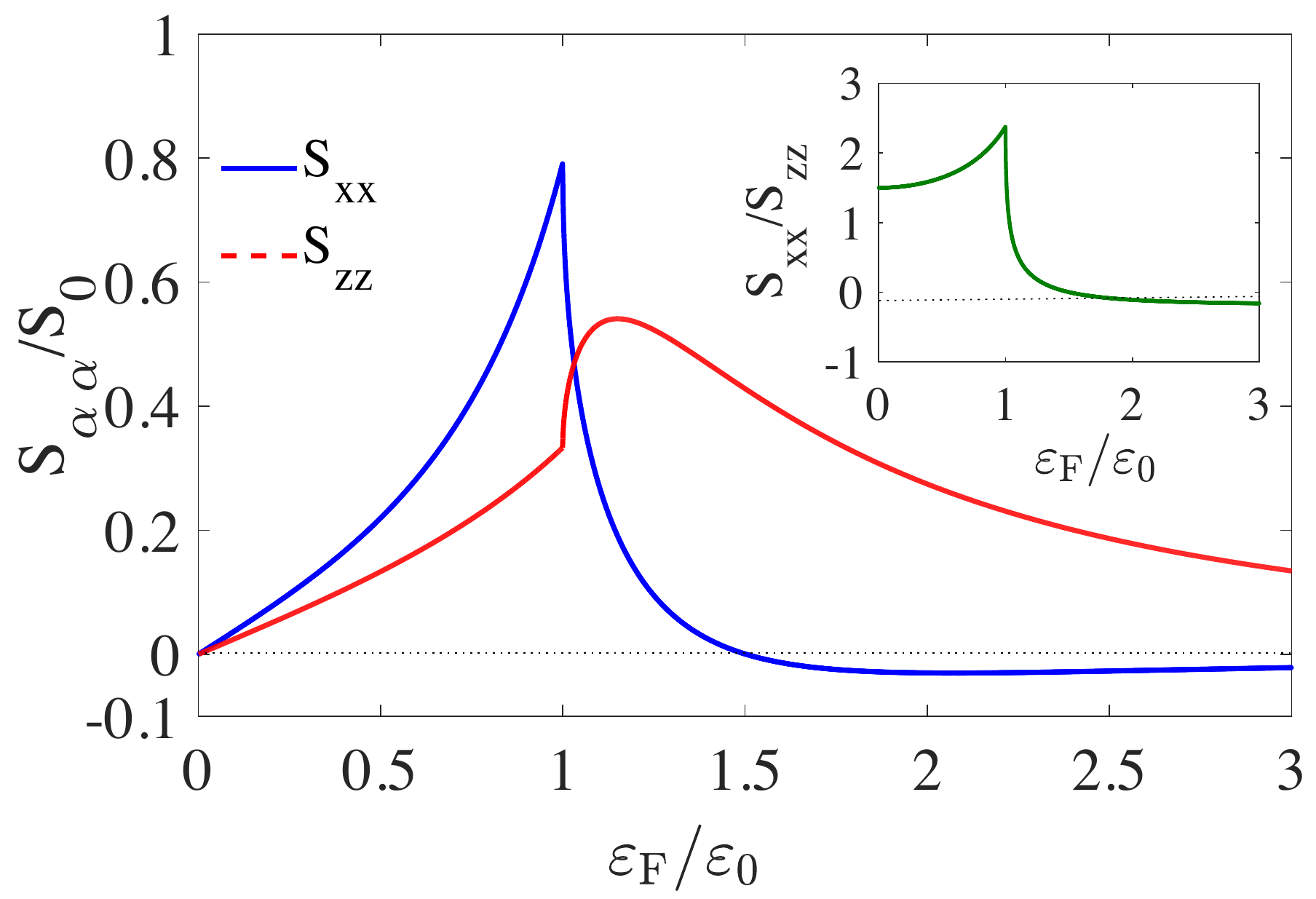}
\caption{
The Fermi energy dependence of the radial (solid blue) and axial (dashed red) thermopowers (in units of $S_0$) obtained from the anisotropic relaxation time. 
Inset: The ratio between radial and axial thermopowers $S_{xx}/S_{zz}$ versus the Fermi energy. 
\label{fig:seebeck}}
\end{figure}

The low temperature thermopower along different directions, could be obtained from the Mott formula Eq.~\eqref{eq:Mott}. The full analytic expression turns out to be very cumbersome, but at low energy regime (i.e., $\varepsilon_{\rm F}<\varepsilon_0$) the results simplify to
\be
\begin{split}
S_{xx}(\varepsilon_{\rm F}<\varepsilon_0)&=8S_0
\frac{1-(8-3{\tilde \varepsilon_{\rm F}}^2)/(4-{\tilde \varepsilon_{\rm F}}^2)^{3/2}}{{\tilde \varepsilon_{\rm F}}^3+4 {\tilde \varepsilon_{\rm F}}\left(1-2/\sqrt{4-{\tilde \varepsilon_{\rm F}}^2}\right)},\\
S_{zz}(\varepsilon_{\rm F}<\varepsilon_0)&=S_0\frac{\tilde \varepsilon_{\rm F}}{4-{\tilde \varepsilon_{\rm F}}^2},
\end{split}
\ee
with $S_0$ defined right after Eq.~\eqref{eq:s_iso}.
The full Fermi energy dependence of the thermopower along the radial and axial directions, as well as the ratio between them, is shown in Fig.~\ref{fig:seebeck}.
It is interesting to notice that the sign of thermopower at low doping is reversed in comparison to the results obtained from the constant relaxation time approximation. Moreover, it becomes anisotropic also at low carrier concentration, even so, the results do not depend explicitly on the ratio between two Fermi velocities $\lambda$.  Another interesting observation is the sharp variation of thermopower in both directions around the nodal ring energy $\varepsilon_{\rm F} \sim \varepsilon_0$. The radial thermopower vanishes and then changes sign at $\varepsilon_{\rm F} \approx 1.5\,\varepsilon_0$.

\section{Summary and Discussion}\label{sec:summary}
Using the semiclassical Boltzmann transport theory and the relaxation time approximation we have studied the low-temperature thermoelectric properties of a three-dimensional nodal-line semimetal. As the low energy electron structure of this system is strongly anisotropic, even isotropic impurity potential would make the scattering and therefore the relaxation times anisotropic. 
With proper treatment of this anisotropy for short-range scattering, we obtain fully analytic results for the relaxation times, electrical conductivities, and thermopower.
Thermoelectric responses have strong energy and directional dependence. In particular in the transition from low to high carrier concentration regimes, where the Fermi surface undergoes a topological transition from a torus-like shape into an ellipsoid, a very sharp change in the thermopower is noticeable. 
This suggests the measurement of thermoelectric properties as a simple probe of the topology of the Fermi surface.
Observation of the thermoelectric performance enhancement across the Lifshitz transition has been reported recently~\cite{nishimura_prl2019}. 
We have also investigated the thermoelectric responses obtained within the constant relaxation time approximation, a widely adopted scheme for studying real materials. It is evident that such a simple approximation completely fails in capturing the true behavior of the system at low temperatures. 

In order to get a rough estimation of what we have obtained here, for real materials~\cite{xu_prb2017}, if we take $\varepsilon_0\approx  0.1 \, {\rm eV}$ as the energy of the nodal ring, we find $S_0\approx 70 \, \mu {\rm V}/{\rm K}$ at room temperature. Moreover, taking $k_0 \approx 0.2\,\pi/a$ for the nodal ring radius, where $a$ is the size of a lattice primitive vector and can be chosen to be $a \approx 5\, \AA$, using $\lambda =1$ and $g=2$ we find $n_0 \approx 3 \times 10^{20} \,{\rm cm}^{-3}$ for the density of electrons in the conduction band corresponding to $\varepsilon_{\rm F}=\varepsilon_0$. This is comparable with the carrier concentration in heavily doped semiconductors.

Finally, we should note that the real potential of nodal-line semimetals for thermoelectric applications requires a more thorough investigation of their thermoelectric behavior in higher temperatures and with the inclusion of different scattering mechanisms, as well as the phononic contribution to their thermal conductivity.

\acknowledgments{We thank Habib Rostami and Keivan Esfarjani for helpful discussions. This work is supported by Iran Science Elites Federation (ISEF).}

\appendix
\section{The density of states, carrier concentration, and chemical potential}\label{app:chemical}
The density-of-states per unit volume of system is given by
\be\label{eq:dos}
\begin{split}
\rho(\varepsilon)&=\frac{g}{V}\sum_{\kv,s}\delta(\varepsilon-\varepsilon_{\kv,s})\\
&=\rho_0 |\tilde \varepsilon|
\left\{1+\frac{\Theta(|\tilde \varepsilon|-1)}{\pi}
\left[\sqrt{\tilde \varepsilon^2-1}-{\rm arcsec}(|\tilde \varepsilon|)\right]\right\},
\end{split}
\ee
with $\rho_0=g k_0^2/(2\pi\lambda \hbar v)$ and $\tilde \varepsilon=\varepsilon/\varepsilon_0$.
If we consider an electron doped system (i.e., $\varepsilon_{\rm F}>0$), the carrier concentration in the conduction band could be readily obtained as
\be
\begin{split}
n_c&=\int_0^{\varepsilon_{\rm F}}\mathrm{d}\varepsilon\, \rho(\varepsilon)\\
&=n_0 {\tilde \varepsilon_{\rm F}}^2
\left\{1+\frac{\Theta({\tilde \varepsilon_{\rm F}}-1)}{\pi}
\left[\frac{2+\tilde \varepsilon_{\rm F}^{-2}}{3}\sqrt{\tilde \varepsilon_{\rm F}^2-1}-{\rm arcsec}({\tilde \varepsilon_{\rm F}})\right]\right\},
\end{split}
\ee
where $n_0=\varepsilon_0\rho_0/2$ is the density of electrons in the conduction band corresponding to $\varepsilon_{\rm F}=\varepsilon_0$.

The chemical potential $\mu$ generally depends on the temperature.  
At low temperatures we have ~\cite{gorsso_book}
\be
\dfrac{d\mu}{dT}=-\dfrac{\pi^2}{3}k_B^2T\dfrac{\rho'(\mu)}{\rho(\mu)},
\ee
where $\rho'(\varepsilon)$ is the derivative of the density-of-states given by Eq.~\eqref{eq:dos}.
Now, using the fact that $\mu(T=0)=\varepsilon_{\rm F}$, we find
\be
\mu(T)\approx \varepsilon_{\rm F}\left[1-\alpha(\tilde \varepsilon_{\rm F})\frac{\pi^2}{6}\left(\frac{k_{\rm B}T}{\varepsilon_{\rm F}}\right)^2\right],
 \ee
with
\be
\alpha(x)=1+\Theta(x-1)\frac{\sqrt{x^2-1}}{\pi+\sqrt{x^2-1}-{\rm arcsec}(x)}.
\ee

\section{Calculation of the anisotropic relaxation times}\label{app:relaxation}
In this section, we provide the details of obtaining the anisotropic relaxation times $\tau^{\alpha}_\kv$. 
As we have explained in the main text, the relaxation times of an anisotropic system could be obtained from the solution of the integral equation~\eqref{eq:relaxtimeaniso}, which for our nodal-line system simplifies to
\be\label{eq:tau_ab} 
 \begin{aligned}
1=\tilde{\tau}^{\alpha}_{\kv}\left(b_0+b_c \cos\theta\right)
-\frac{1}{{\tilde v}^{\alpha}_{\kv}}
\left[a_0^\alpha+a_c^\alpha\cos\theta+a_s^\alpha\sin\theta\right]
\end{aligned}
\ee
where ${\tilde v}^{\alpha}_{\kv}={v}^{\alpha}_{\kv}/v$, $\tilde{\tau}^{\alpha}_\kv={\tilde \kappa} \tau^{\alpha}_\kv/\tau_0$
with $\tau_0=4\pi \lambda /(u_0k_0^2)$ and ${\tilde \kappa}=\kappa/k_0$. 
The dimensionless coefficients $b_i$ are given by
\be
\begin{split}
b_0({\tilde \kappa})&\equiv\int\frac{\mathrm{d}\theta}{\pi}(1+{\tilde \kappa}\cos\theta)\\
&=2\left\{1+\frac{\Theta(\tilde \kappa-1)}{\pi}\left[\sqrt{{\tilde \kappa}^2-1}-{\rm arcsec}(\tilde \kappa)\right]\right\},
\end{split}
\ee
and
\be
\begin{split}
b_c({\tilde \kappa})&\equiv\int\frac{\mathrm{d}\theta}{\pi}(1+{\tilde \kappa}\cos\theta)\cos\theta\\
&={\tilde \kappa}\left\{1+\frac{\Theta(\tilde \kappa-1)}{\pi}\left[\frac{\sqrt{{\tilde \kappa}^2-1}}{\tilde \kappa^2}-{\rm arcsec}(\tilde \kappa)\right]\right\}.
\end{split}
\ee
Furthermore, the dimensionless coefficients $a_i^\alpha$ are defined in terms of the relaxation times
\be\label{eq:a_0cs}
\begin{split}
&a_0^\alpha({\tilde \kappa})\equiv \int\frac{\mathrm{d}\theta'\mathrm{d}\phi'}{2\pi^2}(1+{\tilde \kappa}\cos\theta')
{\tilde \tau}^\alpha_{\kappa'}{\tilde v}^\alpha_{\kappa'},\\
&a_c^\alpha({\tilde \kappa})\equiv \int\frac{\mathrm{d}\theta'\mathrm{d}\phi'}{2\pi^2}(1+{\tilde \kappa}\cos\theta')\cos\theta'
{\tilde \tau}^\alpha_{\kappa'}{\tilde v}^\alpha_{\kappa'},\\
&a_s^\alpha({\tilde \kappa})\equiv \int\frac{\mathrm{d}\theta'\mathrm{d}\phi'}{2\pi^2}(1+{\tilde \kappa}\cos\theta')\sin\theta'
{\tilde \tau}^\alpha_{\kappa'}{\tilde v}^\alpha_{\kappa'}.
\end{split}
\ee
Now, substituting the group velocities from Eq.~\eqref{eq:velocity} into Eq.~\eqref{eq:tau_ab}, it is easy to realize that
\be\label{eq:tau_tilde_xz}
\begin{split}
\tilde{\tau}^x_\kv\left(b_0+b_c\cos\theta\right)
&=1+\frac{a^x_0+a^x_c\cos\theta+a^x_s\sin\theta}{\cos\theta\cos\phi},\\
\tilde{\tau}^z_\kv\left(b_0+b_c\cos\theta\right)
&=1+a^z_s+\frac{a^z_0+a^z_c\cos\theta}{\sin\theta},
\end{split}
\ee
where the ${\tilde \kappa}$ dependance of the coefficients have been dropped for convenience.
The coefficients $a_i^\alpha$ could be obtained after replacing the expressions for the relaxation times from Eq.~\eqref{eq:tau_tilde_xz} back into Eq.~\eqref{eq:a_0cs}. However, with a simple inspection, it becomes clear that the last terms on the right-hand-sides of Eq.~\eqref{eq:tau_tilde_xz} do not survive the angular integrations in Eq.~\eqref{eq:sigma} and therefore do not have any contribution to the thermoelectric responses. 
The only coefficient which needs to be determined is $a_s^z$, which reads
\be
a^z_s({\tilde \kappa})=\frac{\gamma_s({\tilde \kappa})}{1-\gamma_s({\tilde \kappa})},
\ee
with
\be
\begin{split}
\gamma_s({\tilde \kappa})&\equiv
\int\frac{\mathrm{d}\theta}{\pi}\frac{1+{\tilde \kappa}\cos\theta}{b_0+b_c\cos\theta}\sin^2\theta\\
&=\frac{{\tilde \kappa}}{\pi b_c}\left(\pi-\theta_0+\sin\theta_0\cos\theta_0\right)\\
&~~~~+\frac{2(b_0{\tilde \kappa}-b_c)}{\pi b_c^3} \Bigg[b_c\sin\theta_0-b_0(\pi-\theta_0)\\
&~~~~~~~~~~~+2\sqrt{b_0^2-b_c^2} \arctan\left(\sqrt{\frac{b_0-b_c}{b_0+b_c}}\cot{\frac{\theta_0}{2}}\right)
\Bigg].
\end{split}
\ee
Furthermore, in a similar fashion we can define
\be
\begin{split}
\gamma_c({\tilde \kappa})\equiv&
\int\frac{\mathrm{d}\theta}{\pi}\frac{1+{\tilde \kappa}\cos\theta}{b_0+b_c\cos\theta}\cos^2\theta \\
=&-\gamma_s(\tilde \kappa)+\frac{2{\tilde \kappa}}{\pi b_c}(\pi-\theta_0)\\
&-\frac{4(b_0{\tilde \kappa}-b_c)}{\pi b_c\sqrt{b_0^2-b_c^2}}\arctan\left(\sqrt{\frac{b_0-b_c}{b_0+b_c}}\cot{\frac{\theta_0}{2}}\right),
\end{split} 
\ee
which is used in the main text to express $\sigma_{xx}$.
Now, we define the modified relaxation times 
\be
\begin{split}
\delta \tilde{\tau}^x_\kv
&=\frac{1}{b_0({\tilde \kappa})+b_c({\tilde \kappa})\cos\theta},\\
\delta \tilde{\tau}^z_\kv
&=\frac{1}{b_0({\tilde \kappa})+b_c({\tilde \kappa})\cos\theta}\frac{1}{1-\gamma_s({\tilde \kappa})}.
\end{split}
\ee
which are simply the parts of $\tau^\alpha_\kv$, contributing to the thermoelectric responses.


\end{document}